\documentstyle[rotate,psfig,dina4,12pt]{article}
\newcommand{\be}{\begin{equation}}
\newcommand{\ee}{\end{equation}}
\newcommand{\bea}{\begin{eqnarray}}
\newcommand{\eea}{\end{eqnarray}}

\begin{document}
\title{%
Dynamical properties of flux tubes in the
Friedberg-Lee model}
%\thanks{Work supported by BMBF and GSI Darmstadt.}}
\author{%
S. Loh,
C. Greiner\footnotemark[2], M.H. Thoma and U. Mosel\\
Institut f\"ur Theoretische Physik, Universit\"at Giessen\\
D-35392 Giessen, Germany\\%
}
\date{}
\maketitle
\footnotetext[2]{Contribution to the XXXV International Winter Meeting on Nuclear
Physics, Bormio (Italy), \newline 3 - 8 Feb. 1997}
\vspace{-1cm}
\begin{abstract}
A dynamical model of confinement based on a
microscopic transport
description of the Friedberg-Lee model
is extended to explicit color degrees of freedom. The string tension is
reproduced by an adiabatic string formation from
the nucleon ground state.
As a particular application,
we address the question of how a charmonium state might
be dissociated by the strong color electric fields
when moving through a color electric flux tube and
speculate on the importance of such an effect
with respect to the issue of $J/\psi $--suppression observed in ultrarelativistic
heavy ion collisions.
Furthermore,
we show the dynamical breakup
of flux tubes via $q-\bar{q}$-particle production and the disintegration
into mesons. There we encounter some problems within the Vlasov-type
realization of describing the motion of the quarks which can be resolved
by a molecular dynamical approach.
\end{abstract}

\section{\bf Introduction}

\bigskip
One of the main goals of ultrarelativistic heavy ion physics is the
possible formation of a deconfined state of hot and dense nuclear matter,
the quark-gluon plasma, as it is believed that the energy density reached
may be high enough for its formation. On the other side,
in popular microscopic models, simulating the whole reaction
of the collision, strings are produced in the first few moments of the
interaction, which subsequently decay (fragment) into secondaries
(mesons, baryons, string-like hadronic resonances). These are best visualized
as color electric flux tubes. Their creation is thought to happen by the exchange of a
color octet gluon or by colorless momentum transfer
among target and projectile nucleons. (The color sources in the endcaps
of the flux-tubes should be thought as a quark on the one side and an
anti-quark on the other side, or as a quark and a di-quark, respectively.)
Before a further hadronic-like or partonic-like state of matter forms, such
a temporary build-up and decay of strings is
assumed to describe the very early collision phase.
A serious problem of these models is the
description of the intermediate state of the reaction, the hadronization
of the ensemble of strings, which can not be treated perturbatively,
since the hadrons are formed by the purely nonperturbative effects of
color confinement. Therefore the dynamics of confinement should be
considered in the transport descriptions.

We describe the dynamics of a flux tube within the
nontopological soliton model of {\em Friedberg} and {\em Lee}
\cite{FL77,Wilbuch}
supplemented by a semiclassical transport equation for the quarks.
In this effective model
the nonperturbative effects are modeled by the presence
of a scalar field. Since in this model the hadron surface is generated
dynamically, it is well suited for dynamical simulations. Furthermore
absolute confinement is implemented by the idea of
describing the vacuum as a color dielectric medium. In its original version
the Lagrangian is given by
\be
  \label{lagrange}
  {\cal{L}} = \bar{\Psi}(i\gamma_{\mu}\partial^{\mu}-m_0-g_0\sigma)\Psi
              + {1 \over 2}(\partial_{\mu}\sigma)^2 - U(\sigma) 
              - {1 \over 4} \kappa(\sigma)F_{\mu\nu}^aF^{\mu\nu}_a 
         - ig_v\bar{\Psi}\gamma_{\mu}{\lambda^a \over 2}\Psi A^{\mu}_a \; ,
\ee
where $\Psi$ denotes the quark fields, $\sigma$ is the color singlet scalar
field representing the long range and nonabelian effects (of multi gluon
exchange). The last term contains the interaction of the residual
classical and abelian color fields $A_{\mu}^a$.
Color confinement is obtained by the general properties of $\kappa$.
Nonabelian effects are assumed to be absorbed in the color
dielectric function $\kappa(\sigma)$ which is chosen such that
$\kappa$ vanishes as $\sigma$ approaches its vacuum value $\sigma = \sigma_v$
outside the bag and $\kappa = 1$ inside,
\be
  \kappa(\sigma) = |1 - {\sigma \over \sigma_{vac}}|^2
                   \Theta(\sigma_{vac} - \sigma) \; .
\ee
\be
 U(\sigma) = {a \over 2!}\sigma^2 + {b \over 3!}\sigma^3 +     
             {c \over 4!}\sigma^4 + B    
\ee
is the self-interaction potential for the scalar
$\sigma$-field containing cubic and quartic terms.

In section II we present the resulting semiclassical transport equations
\cite{LoBiMoTh96,Lo96a}. In section III we summarize how the parameters
of the model can be fixed by static properties of hadrons and flux tubes.
In section IV we adress the question what happens if
a $J/\psi $--state moves into an environment
of color electric strings and the implications of this investigation
on the issue of $J/\psi $--suppression.
Finally we present in section V
the full transport dynamical approach to the disintegration
of a flux tube into mesons via quark antiquark particle production. After the
discussion of some intrinsic problems connected with the semiclassical limit of
the model, we give the space-time description of a string-fragmentation process
of the Lund-type within a molecular dynamical approach.

\section{\bf Transport equations}

The quark degrees
of freedom are effectively handled via a transport equation derived by Elze and Heinz
\cite{ElHe89} in the semiclassical limit.
Since the gluon field is effectively abelian (i.e. a Maxwell field),
one can work in a U(1) sub-space of SU(3)-color.
One obtains \cite{LoBiMoTh96,Lo96a}
two transport equations for the
phase space distributions of
color charges, i.e. quarks ($f$) and antiquarks (or diquarks) ($\bar{f}$)
\bea
 & & (p_{\mu}\partial^{\mu}-m^*(\partial_{\mu}m^*)\partial^{\mu}_p)f(x,p) =
  g_vp_{\mu}F^{\mu\nu}\partial_{\nu}^pf(x,p)  \nonumber \\
  \label{vlas2}
 & & (p_{\mu}\partial^{\mu}-m^*(\partial_{\mu}m^*)\partial^{\mu}_p)\bar{f}(x,p) =
  -g_vp_{\mu}F^{\mu\nu}\partial_{\nu}^p\bar{f}(x,p) \; ,
\eea
which is a set of usual Vlasov equations describing the motion of charged
particles in a selfconsistently generated
mean scalar and vector field
determined below ($m^*=m_0+g_0 \langle \sigma \rangle $).
Within this description, the coupling of quark and antiquark degrees of
freedom is provided by their interaction with the mean fields $\sigma$ and
the colorelectric field only.
The equations are are solved numerically by applying the test-paticle method.

The $\sigma$ and the $A_{\mu}$-field
are treated as classical fields (mean field approximation).
The equation of motion for the scalar soliton field then reads
\be
  \label{klegor}
  \partial_{\mu}\partial^{\mu}\sigma + U'(\sigma)
  + \frac{1}{4} \kappa'(\sigma)
  F_{\mu\nu}F^{\mu\nu} + g_0\bar{\Psi}\Psi = 0
  \;,
\ee
which is solved using a
staggered leapfrog algorithm
\cite{Vet95}.
For the gluon field
we choose the Coulomb gauge $\vec{\nabla}\cdot(\kappa\vec{A}) = 0$ resulting in
\bea
  \label{maxsca}
  \vec{\nabla}(\kappa\vec{\nabla}A_0) &=& -j_0 \; , \\
  \label{maxvec}
  -\kappa\partial_t^2 \vec{A} + \vec{\nabla}^2\kappa \vec{A} -
  \vec{\nabla}\times (\kappa \vec{A} \times {\vec{\nabla}\kappa \over \kappa})
  &=& - \vec{j} + \kappa\vec{\nabla}\partial_tA_0 \; .
\eea
In the following equation (\ref{maxvec}) is neglected as it can be shown, that
currents within a flux tube do not produce a magnetic field because the
displacement current is exactly cancelled by the convection current
if the string radius stays nearly constant \cite{WiPu95}
(and which is exact in pure 1+1-dimensional
electrodynamics).
As a consequence,
we determine the colorelectric field instantaneously by
$\vec{E} = -\vec{\nabla}A_0$,
which is determined by a
two-dimensional finite element method
(appropriate for cylinder symmetrical configurations).

The color
charge density entering into (\ref{maxvec}) is expressed by
\be
  \label{coldens}
   j_0 = {\eta \over (2\pi)^3} \int d^3p (f(x,p)-\bar{f}(x,p)) \; ,
\ee
whereas the scalar density $\rho_s \equiv \bar{\Psi } \Psi $ entering into
eq. (\ref{klegor}) reads
\be
  \label{sdens}
  \rho_s = {\eta \over (2\pi)^3} \int d^3p {m^* \over \omega}
           (f(x,p)+\bar{f}(x,p)) \; .
\ee
($\eta \equiv 4$ accounts for spin and flavour degeneracy.)

\section{Static properties}

We first examine the static limit of the transport equations.
The color charge density in a hadron has to vanish
locally, i.e. $j_0(x) \equiv 0$.
From that one concludes that
there is no colorelectric field in the groundstate, thus $f = \bar{f}$.
Because of the mean field approximation for the scalar field
the distribution functions have to be
of a local Thomas Fermi type
\be
  \label{thofer}
  f(x,p) = \bar{f}(x,p) = \Theta (\mu-\omega) \; ,
\ee
where we have introduced the Fermi energy $\mu$ for the quarks.
Considering these groundstate distributions, the quark degrees can be
integrated out resulting
in a selfconsistency relation for the $\sigma $-field in eq. (\ref{klegor}).
The soliton solution is con\-struc\-ted using a shoot\-ing me\-thod of
Van Wijn\-gaar\-den-Dekker-Brent \cite{Numerical}.

The model parameters are now adjusted
in a way to reproduce the quark-number, mean mass of delta
and nucleon and the rms radius of the nucleon for a typical baryon. With
these parameters (see table \ref{tab1}) we use then a new Fermi energy $\mu$ in order
to go from the baryonic solution ($N_Q=3$),
resembling a quark-diquark configuration to the mesonic solution ($N_Q=2$),
resembling a quark-antiquark configuration.
The physical properties of these soliton
solutions are shown in table \ref{tab2} and \ref{tab3}, where the meson mass,
which typically is too high in these models, can be significantly reduced by
the inclusion of momentum projection methods and the colormagnetic energies
\cite{Wilbuch}.
The remaining properties are very well reproduced within
our model approach. These groundstate solutions will be used in the following
sections for dynamical applications.
\begin{table} [ht]
\center{\begin{tabular}{||c|c||} \hline
parameter set &   \\ \hline
$a$ $[fm^{-2}]$   & 0.0      \\
$b$ $[fm^{-1}]$   & -419.3   \\
$c$ $[1]$         & 4973.0       \\
$B$ $[fm^{-4}]$   & 0.283      \\
$g_0$ $[1]$       & 8.0         \\
$m_0$ $[fm^{-1}]$ & 0.025       \\
$\mu$ $[fm^{-1}]$ & 1.768 / 1.9582 \\
$\eta$            & 4         \\
$\sigma_{vac}$ $[fm^{-1}]$ & 0.253 \\ \hline
glueball mass [$GeV$] & 1.045\\
\hline
\end{tabular}}
\caption{The table shows the parameters used to construct a mesonic and
a baryonic soliton solution of the sigma field equation. The first value of
$\mu$ is used for the baryonic solution, the second one for the mesonic.}
\label{tab1}
\end{table}
\begin{table} [ht]
\center{\begin{tabular}{||c|c|c||} \hline
Parameter set & meson & Experimental Data \\ \hline
%$Quark number$$[1]$& 1.0        &  1.0  \\
$E$ $[MeV]$        & 798        &  465  \\
$RMS$ $[fm]$       & 0.653      &  0.66 \\
\hline
\end{tabular}}
\caption{The table shows the results
of the fits for the meson compared to the experimental data.
The experimental data are the mean values of the pion and the rho-meson
(taken from \protect\cite{Wilbuch}).}
\label{tab2}
\end{table}
\begin{table} [ht]
\center{\begin{tabular}{||c|c|c||} \hline
Parameter set & baryon & Experimental Data \\ \hline
%$Quark number$$[1]$& 1.0        &  1.0  \\
$E$ $[MeV]$        & 1099       &  1087 \\
$RMS$ $[fm]$       & 0.693      &  0.83 \\
\hline
\end{tabular}}
\caption{The table shows the results of the fit for the baryon compared to the
experimental data. The experimental data are taken from \protect\cite{Wilbuch}.}
\label{tab3}
\end{table}

Since the colorelectric field vanishes in the static solution,
the strong coupling constant $\alpha_s = g_v^2 / 4 \pi$
cannot be fixed from ground state properties. In order to determine its value
we investigate the colorelectric energy in the cavity by forming a string
like configuration (see fig. 1).
We find
the typical string like behaviour of the $\vec{E}$-field \cite{LoBiMoTh96}, namely
being constant and almost independent of the axial coordinate in the region
between the quarks.
The genuine feature of this field is, that it is parallel everywhere at
the boundary of the cavity, which is due to
the von Neumann boundary conditions, forcing the normal component of the
displacement field $\vec{D} = \kappa \vec{E}$ to vanish.
Determining the colorelectric energy $E_{glue} =
{1 \over 2} \int d^3r \vec{D} \vec{E}$ and the energy carried by the
scalar field
as a function of the length of the string provides
us with a measure for the string constant $\tau$.
In order to obtain the generally accepted value of $\tau \approx 1 GeV/fm$,
we need to have values of $\alpha_s \approx 2$, in accordance with other
estimates \cite{Wilbuch}.
The volume averaged electric field
$\langle E_z \rangle $ takes a value of 1.6 fm$^{-2}$
with $E_{z,max} =2.2$ fm$^{-2}$ along the z-axis. The radius
of the string results as approximately 0.8 fm.

%
%\begin{figure}
%\vspace*{4cm}
%{\small
%Figure 1: The electric field {\bf E} inside a cavity (flux tube) of
%approximately 12 fm in length.}
%\end{figure}

One can also extract a string-string interaction potential by an
adiabatic fusion of two parallel or antiparallel strings \cite{Lo96a}.
The potential
is obtained by comparing the colorelectric and the $\sigma-$field energies
of the fused configuration to the (infinitely) separated configuration.
Due to confinement in this model there is no long range interaction (over
distances typically of the order of $1\, fm$ and more).
In the case of two parallel strings,
in the intermediate
range, when the confinement wall starts to break down, we have a balance
of gaining surface energy from the $\sigma-$field and increasing the total
energy from the coherent addition of the colorelectric field.
In the short range region the potential is strongly repulsive due to
the unscreened quark color charges at the endcaps.
%In the case of antiparallel strings, however, a strong attraction is
%found leading to a neutralization of the color charges and the collapse of
%the flux tube.

\section{$J/\psi$-dissociation by a color flux tube}

Since the original work of Matsui and Satz
\cite{Mat86} who proposed $J/\psi $--suppression
as a clear signal for quark gluon plasma formation (the electric forces responsible
for the binding are sufficiently screened by the plasma so that no boundstate
should form) an intense work by several experimental groups
(NA38-collaboration) has adressed this
issue \cite{Exp}. Indeed a suppression in the corresponding dilepton signal
compared to the Drell-Yan background has been seen over the last years
for lighter projectiles like O+Cu, O+U and S+U. However,
these observations may also be explained alternatively:
(1) A part or even the total suppression can probably be explained
by $J/\psi $--absorption on the surrounding nucleons
\cite{Ger92},
i.e. $J/\psi + N \rightarrow \Lambda_C + \bar{D}$;
(2) additional absorption
by exothermic reactions like
i.e. $J/\psi + \rho \rightarrow D + \bar{D}$
might be attributed to `comovers' (`mesons')
being produced as secondaries.
Indeed, if $\sigma _{abs}^{\psi N} \approx 7$ mb is assumed,
the reported
suppression could be nicely reproduced by the absorption within the nuclear
environment.
In any case, the general consensus has been, that a highly (energy) dense
intermediate reaction zone is needed to explain the observed suppression.
Recently the data taken in Pb+Pb reactions at 160 AGeV
by the NA50-collaboration \cite{Exp1} have given new excitement
to the whole issue as a stronger absorption has been found than suggested
by the models of absorption models on nucleons.

The $J/\psi $ (or better $c\bar{c}$-state which later will form the $J/\psi $)
as a rather heavy hadronic particle is being produced at the earliest state of the reaction
in a hard collision among the nucleons. Thus it is natural to ask what happens if
a (pre-) $J/\psi $--state (as a strongly interacting probe)
enters into an environment of color electric strings.
As the string carries a lot of internal energy (to produce the later secondaries)
the quarkonia state might get absorbed and completely dissociated by the
intense color electric field inside a single flux tube.

To adress this question,
we first describe semiclassically
a bound $c\bar{c}$-state (synonymously a $J/\psi $) and also
a bound $c\bar{q}$-state (synonymously a $D$) within our model.
We start from a light meson state
and then adiabatically
increase the mass of the initial light quark(s)
($m_{i,q} \approx 10$ MeV) to a final heavy quark mass
(for details see \cite{Lo97}).
Taking the a priori unknown charm quark mass as $\approx \, 1.25$ GeV,
the overall mass of the states are found as
\bea
   \label{mass}
     m_{c\bar{c}} & \approx & 3 \, \mbox{GeV} \\
     m_{c\bar{q}} & \approx & 1.7 \, \mbox{GeV} \, \, \, .\nonumber
\eea
The mass of the $c\bar{c}$-state lies only slightly above the experimental
value by about $\approx 60$ MeV, whereas the mass of the $c\bar{q}$-state
is lowered by about the same value, thus providing a very good description.
The radius of the heavy meson
states (the `$J/\psi $'- and `$D$'-meson) are in the right range of what
one would expect from other quarkonia model.

We now insert the `$J/\psi $' `by hand' into
the central interior of a stationary flux tube (see figs. 2).
Initializing
a cylindrically symmetric configuration in this way, we proceed by solving
the full set of dynamical equations of motion, i.e. the two
Vlasov equations (\ref{vlas2}) for the heavy quark and antiquark distribution,
the equation (\ref{klegor}) for the
soliton field $\sigma $ and the one for the electric potential (\ref{maxsca}).
What happens is that the
heavy quark is steadily
pulled to one endcap of the string carrying opposite charge,
whereas the antiquark correspondingly moves
into the other direction. Being pulled apart, the two
displaced charges screen the overall electric field in between.
In other words, the electric field energy originally stored
is transformed into the kinetic energy of the oppositely moving
two heavy quarks.
In figs. 2 also the evolution of the $\sigma $-field is depicted
for various time stages. At the initial times one clearly sees how
the soliton solution of the flux tube is distorted by introducing
the $c\bar{c}$-state into the center of the string.
At the timesteps up to about 5 fm/c the $\sigma $-field follows the
quark distribution quite well, its shape reflects the shape of the
latter distribution.
In the
further evolution (not shown) the soliton field relaxes much more slowly to its
vacuum value compared to the more or less instantanous screening
of the electric field. It typically takes about 5 fm/c until $\sigma $
turns for the first time to its vacuum value.
This long time interval
is basically a consequence of the nonlinear potential $U(\sigma )$
being flat around $\sigma \approx 0$ and corresponds to the time
needed for the soliton field to `move down' the potential hill to
its vacuum value at $\sigma = \sigma_V$.
The original field
energy carried by the soliton field (the `Bag' energy) being still present
after the heavy quarks have already moved apart is initially more or less
conserved. The late oscillations observed in the simulations are
only damped by a transverse expansion of the field itself, which,
however, is a rather slow process because of the large `glueball' mass
of the field.
If one would consider a chiral O(4)- extension
for the soliton field one expects that these oscillations
might accordingly transform into low-momentum (and nearly massless)
pion modes and should accordingly be damped away more quickly.

%\begin{figure}
%\vspace*{8cm}
%{\small
%Figure 2:
%The quark distribution (left side) and the scalar field
%distribution (right side) at various times in the evolution
%($t=0 \, 0.8, \, 3.4, \, 4.6$ fm/c) are shown in a contourplot.}
%\end{figure}
%
From this simulation we conclude that
a localized $c\bar{c}$-state immediately gets separated by the strong
colorelectric field, and, in return, will finally be disintegrated
into $D$-meson (or $\Lambda _C$-baryon) like configurations.
The dissociation corresponds to classical field ionization.
Such a dissociation should also persist if initially
the $c\bar{c}$-state enters the string with some moderate
or even quite large transverse momentum \cite{Lo97}.

%
%\begin{figure}
%\vspace*{4cm}
%{\small
%Figure 3:
%The position of nucleonic strings is plotted
%in the cms frame in a central S+U reaction at 200AGeV.
%Each string is accompanied by a circle of radius
%0.5 fm.}
%\end{figure}
%

But what is now the actual situation in a relativistic heavy ion collision?
If one believes in the successful descriptions
of microscopic approaches, a large
region in space in the first few moments after the reaction is spanned
by highly excited longitudinal strings.
Especially for the more
heavy systems the effective region (or volume) of all the strings
being produced within a short time interval of less then one fm/c
gets so large that the strings become closely packed or already
overlap. Although most often the strings are thought to fragment
independently one might also consider the possibility
of color rope formation as higher charged tubes.
For a quantitative estimate we depict in fig. 3
the position of nucleonic strings (highly energetic strings excited by
a target and projectile nucleon)
all being produced in a cms-time interval
of $\approx 0.5 $ fm/c within a slab of the complete transverse area
and 1 fm thickness along the longitudinal direction
in the cms frame in a central S+U reaction at 200AGeV generated
within the HSD approach \cite{Ca96}.
This simple reasoning illustrates that
the whole reaction area in an ultrarelativistic heavy ion collisions
might be completely filled by strings in the first few moments.
If such a large spatial region does exist, most
of the produced $J/\psi $'s have to pass it and thus are
affected by this highly excited environment. The suppression of the
$J/\psi$'s will depend crucially on the (average) length of the produced
strings before they hadronize, as this decides about the available `empty'
space for the $J/\psi$'s (or pre-$J/\psi$-states) to escape the initial
reaction zone without entering any of the individual strings being built up.
%If they will get
%dissociated (in part or all of them) this would lead to a
%minor or stronger suppression of $J/\psi $ to be finally observed.
Such a conclusion can only be tested further and strengthened
if one incorporates these ideas within one of the present
microscopic transport algorithms.

\section{Flux tube breaking}

In this section we show how a flux tube of the Friedberg-Lee model breaks up
due to quark-antiquark pair production. One usually describes such processes
with the Schwinger formula, which gives a constant pair
production rate of electron-positron pairs in QED, depending only on the
absolute value of the electric field. However, this formula
cannot be naively transferred to the QCD case of $q\bar{q}-$production in a
flux tube, because
the back reaction of the produced pairs on the external field
has to be considered, since this screening of the field is finally
responsible for the breakup of the strings. In addition one must consider
the modifications of the pair production rate due to transverse and longitudinal
confinement.
A detailed analysis of the time dependent pair production process
within the Friedberg-Lee model has not yet been performed. Therefore we have
to rely on different phenomenological arguments, providing us with a simple
guideline to the space-time evolution of a fragmenting string.
The earliest
and most successful model of string fragmentation is the Lund-model
\cite{AnGuInSj83}. Within this model one assumes the quarks to be massless
and therefore moving with the speed of light. Even the original $Q\bar{Q}-$pair,
generating the string, is supposed to travel on the lightcone. Within
these assumptions it is sufficient to describe the fragmentation process
by a 1+1-dimensional space-time geometry, for it is causally impossible
that the produced $q\bar{q}-$pairs propagate towards the endcaps (the
$Q\bar{Q}-$pair).

Hence, as a prerequisite we assume that the original $Q\bar{Q}-$pair travels
with the speed of light along the $\pm z-$axis, irrespective of the interior
dynamics, which is generated by solving selfconsistently the transport equations
(\ref{vlas2}) together with the mean-field equations for
the $\sigma-$field (\ref{klegor}) and the colorelectric field (\ref{maxsca}).
The produced $q\bar{q}-$pairs
are inserted into the dynamically evolving flux-tube by assuming their groundstate
spatial shape that has been determined in section II and with a
vanishing total momentum \cite{WiPu95}.
If one inserts only one (light) $q\bar{q}-$pair, the fragmentaion of
the strings follow directly to what one would expect from the Lund phenomenology
\cite{Lo96a}. The inserted two light quarks get immediately dissociated
and do follow the two endcaps propagating along the lightcone.
The colorelectric field in between gets completely screened, whereas
the scalar fields is a little retarded in its motion.
From the left part of fig.4 we recognize a small
dispersion of the $q$ and $\bar{q}$ distributions at late times,
which is caused by the
selfinteraction inside of the quark distributions, leading to a screening of the charges:
a charge fragment at the end of one of the two substrings feels a smaller
colorelectric force due to its interaction with the charge fragments in front
of it. This is clearly seen in the right part of fig.~4
where the longitudinal momentum distribution
of the testparticles describing the propagation of the two inserted quarks
is shown. The momentum distribution in $p_z$ of one light quark
is spread from nearly 0 up to 5 GeV/c!

As a further example we show in fig.~5 (left part) the space-time evolution
of the string fragmentation via equal time production of three $q\bar{q}-$pairs
within the full model.
In the upper row at $t=2.8 \, fm/c$,
the string is already extended to a length of $5.6 \, fm$; the
constant colorelectric field along the string is reflected in the constant
field energy between the charges. At $t=3.0\, fm/c$ the three $q\bar{q}-$pairs
are inserted at $z=0\,fm$ and $z=\pm 2.0\, fm$ with a vanishing relative
velocity. In the following time evolution we observe, as in the previous example, the
formation of the outer two meson pairs propagating on the lightcone. In the
inner region of the flux-tube, the respective quark pairs penetrate each other,
having already acquired a small dispersion. At the time $t = 4.2\, fm/c$, the
colorelectric field is almost completely screened. This abrupt change of the
colorelectric field leaves the $\sigma-$field
highly excited. Therefore the $\sigma-$field snaps off and starts to oscillate
around the nonperturbative vacuum value $\sigma_{vac}$.
Thus the $\sigma-$field, which is
supposed to localize the quarks, cannot prevent a further dispersion of the
color charges and the latter neutralize and dissolve along the z-axis
to a length of about $8\, fm$ at $t=8.2\, fm/c$.
We are thus faced with the fact of having
produced real quark-antiquark fragments evaporating in the nonperturbative QCD
vacuum.

%\begin{figure}
%\vspace*{8cm}
%{\small
%Figure 5:  -- Abb. 5.32 aus Doktorarbeit --
%The figure shows the time evolution of the color charge density for
%the breakup of the string via the production of three $q\bar{q}-$pairs
%at different time steps $t=2.6, 3.6, 4.2, 8.2\, fm/c$.
%In the left part
%the motion of the quarks has been calculated within
%the full transport dynamics, whereas
%in the right part the molecular dynamical approach has been used.
%The equidistant
%contour lines run from $-0.5\, fm^{-3}$ to $0.5\, fm^{-3}$.}
%\end{figure}
%

As already mentioned, in the present numerical realization of our model each
testparticle of the respective ensembles interacts not only with the testparticles
of the other quarks, but also with the testparticles of the same quark. This
selfinteraction, which leads to the dispersion of the quark
distributions (see e.g. fig.4), becomes dominant in the limit of only a few interacting
physical particles; in the extreme case, a single color charge distribution
with an almost vanishing mass would immediately dissolve due to the
selfinteraction. Unlike in other models, like e.~g.~the Walecka-model, which
also contains an attractive scalar field and a repulsive vector field and which
has been treated successfully, in the present case the
attractive $\sigma-$field acts only in transverse direction on the produced
quarks. Its longitudinal effect is completely suppressed by the colorelectric
field.
Thus we are forced to formulate a new dynamical approach,
that conserves the identity of the different quarks and antiquarks and is free
of the selfinteraction problem:
A possible way to overcome the problems mentioned above is to treat
the quarks in a molecular dynamical framework \cite{Lo96a}. In this approach each physical
particle (quark or antiquark) is simulated by just one particle and not
by an ensemble of testparticles.
Due to the fact that point-like particles
cause short range divergences in the field equations, a specific spherically
symmetric distribution is assigned to each individual testparticle.
A testparticle then describes the motion of the center of its
distribution according to the Hamiltonian equation of motion.

In the right part of fig.~5 we show the corresponding result within
this molecular dynamical approach.
We see how the screening proceeds not only at the coordinate center, but also
between the other fragments.
In the third and fourth row we clearly
see how the two lightcone fragments have been formed as well as the two excited
'Yo-Yo'-modes.
The confining $\sigma-$field follows the motion of the quarks
with a certain retardation , but finally settles into
individual stable soliton configurations.
In summary, we
find a more or less complete correspondence of the
dynamical behaviour to the one expected from 1+1-dimensional Lund phenomenolgy.

\section{Summary and Outlook}

We have given our present status of a dynamical relization
of the Friedberg-Lee model. This model describes absolute confinement
phenomenologically. The transport equations for the quarks are
of Vlasov-type, whereas the soliton field and the chromoelectric field
are treated as (classical) mean fields. The static properties
of meson and baryons as well as the description of color electric
flux tubes can be succesfully be described within this approach.

As a first dynamical application we have investigated how a $c\bar{c}$-state
(a $J/\psi $) behaves inside a chromoelectric flux tube.
Due to the strong color electric field
inside the flux tube the heavy meson becomes dissociated
(due to the complete screening in effective 1+1-dimensions)
by field ionization rather immediately
(on a timescale of $\stackrel{<}{\approx }1$ fm/c) ending up finally in
$D$- or $\Lambda_c$-like states. It is expected that a large
region in space in the first few moments in
an ultrarelativistic heavy ion collision is spanned
by highly excited longitudinal strings, especially for the more heavy systems.
We have speculated that the dissociation process might provide
an alternative or additional source for explaining
the observed $J/\psi -$suppression.
This possible absorbtion mechanism will be implemented within one of the present
microscopic transport algorithm in order to verify these ideas.

Secondly we have shown how the flux-tube break\-ing pro\-ceeds in the Friedberg-Lee
model.
In the case of three produced
$q\bar{q}-$pairs, the outer meson pairs are formed, but in the center
of the fragmenting string, where the two parallel 'Yo-Yo'-modes are supposed
to build up, we observe a different behaviour:
When the respective quark distributions start to penetrate and neutralize,
a small dispersion of these is already observed. At this time the colorelectric
field is completely screened in the center of the string. This sudden
change of the source of the flux-tube leaves the $\sigma-$field highly excited
and thus it starts to oscillate around the nonperturbative vacuum value.
The $\sigma-$field is not able to prevent the further dispersion
of the quark distributions, which neutralize and dissolve along the
string-axis. With the scalar density being significantly reduced in this case,
the $\sigma-$field undergoes a complete phase transition to the nonperturbative
vacuum.

On the other hand, the dispersion of the quark distributions
is caused by their selfinteraction. This dispersion is caused by the fact,
that we deal with only a few quarks that are treated
as classical charge distributions and interacting strongly via a classical
confining potential. In this case the selfinteraction of the color charges
becomes dominant and generates a screening of the charges, which
finally is responsible for the dispersion of these.
We have shown a possible way to overcome these problems by treating the quarks
and antiquarks as point particles with an appropriately chosen charge distribution
as commonly used in molecular dynamical simulations.

Summarizing we can say that our model provides us with a useful tool to describe
the full dynamical evolution of string formation and decay via multiple
quark-antiquark production. The aim for future investigations should be to
apply our model to more realistic scenarios, like multiple string production
and decay in relativistic heavy-ion collisions, for describing the formation
of hadrons out of an expanding and cooling quark-gluon plasma.

%
%
%\clearpage
%
%\pagestyle{empty}
%
%
\begin{figure}[htbp]
%\vspace{-3cm}
%\centerline{\psfig{figure=c:/users/stefanl/tex/phd/letter/bild2.ps,width=15cm,height=12.5cm}}
%\vspace{-1cm}
{\small
Figure 1: The electric field {\bf E} inside a cavity (flux tube) of
approximately 2 fm and 12 fm in length, respectively.}
\label{efield1}
\end{figure}
%
%\clearpage
%
\begin{figure}
%\vspace{-2cm}
%\centerline{\psfig{figure=c:/users/stefanl/tex/phd/jpsi/fig5.ps,width=15cm}}
%\vspace{-2cm}
{\small
Figure 2:
The quark distribution (left side) and the scalar field
distribution (right side) at various times in the evolution
($t=-0.2, \, 0.8, \, 3.4, \, 4.6$ fm/c) are shown in a contourplot.}
\label{jp1}
\end{figure}
%
%\clearpage
%
\begin{figure}
%\centerline{\psfig{figure=c:/users/stefanl/tex/phd/jpsi/jpfig6.eps,width=12cm}}
%\centerline{\psfig{figure=c:/users/carsten/tex/papers/dissoc/jpfig6.eps,width=12cm}}
{\small
Figure 3:
The position of nucleonic strings is plotted
in the cms frame in a central S+U reaction at 200AGeV.
Each string is accompanied by a circle of radius
0.5 fm.}
\label{figure6}
\end{figure}
%
%\clearpage
%
\begin{figure}
%\vspace*{5cm}
%\centerline{\psfig{figure=c:/users/stefanl/tex/phd/pics/phase1.eps,width=14cm}}
%\vspace{1cm}
{\small
Figure 4:
The left figure shows the color charge density $\rho $
at a late timestage
($\approx 5 \, fm/c$) for the breakup of the a string via the production
of one $q\bar{q}-$pair at time $t=0 \, fm/c$.
The right figure depicts the longitudinal
momentum distribution of the testparticles
of the inserted two light quarks at that time.}
\end{figure}
%
%\clearpage
%
\begin{figure}
%\vspace{-3cm}
%\centerline{\psfig{figure=c:/users/stefanl/tex/phd/pics/compare3.eps,width=15cm}}
%\vspace{-2cm}
{\small
Figure 5: The figure shows the time evolution of the color charge density for
the breakup of the string via the production of three $q\bar{q}-$pairs
at different time steps $t=2.8, 3.6, 4.2, 8.2\, fm/c$.
In the left part
the motion of the quarks has been calculated within
the full transport dynamics, whereas
in the right part the molecular dynamical approach has been used.
The equidistant
contour lines run from $-0.5\, fm^{-3}$ to $0.5\, fm^{-3}$.}
\label{lund4dyn}
\end{figure}
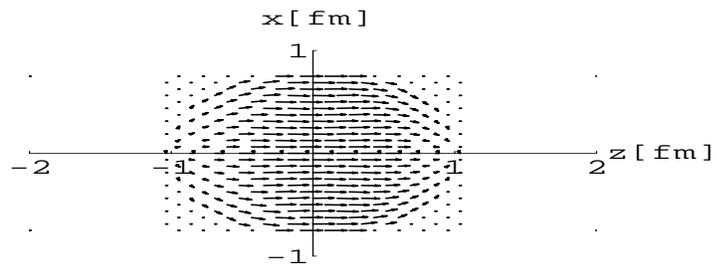
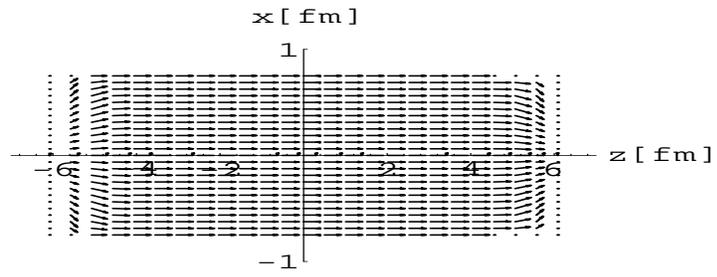
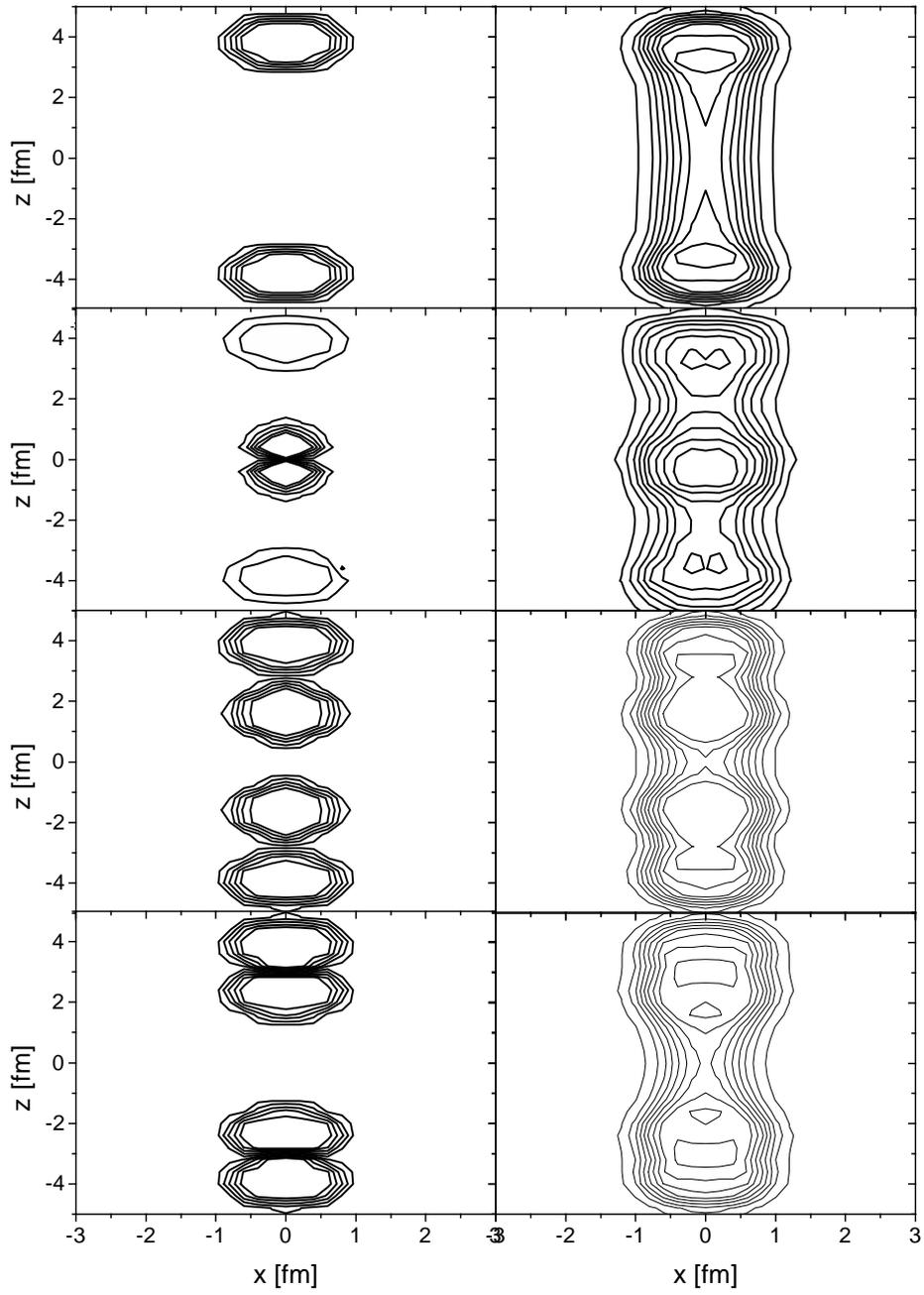
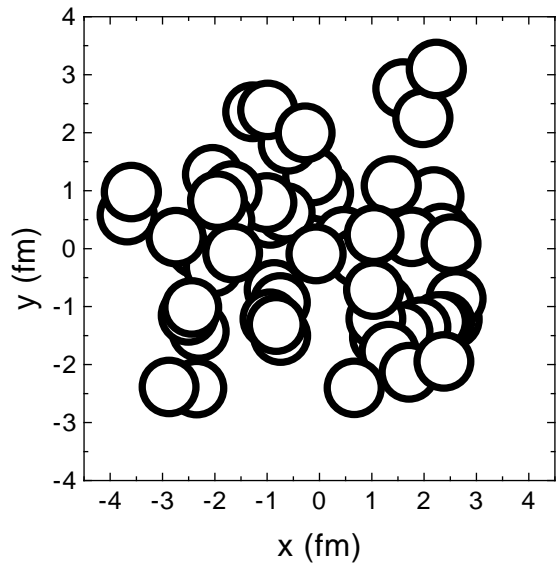
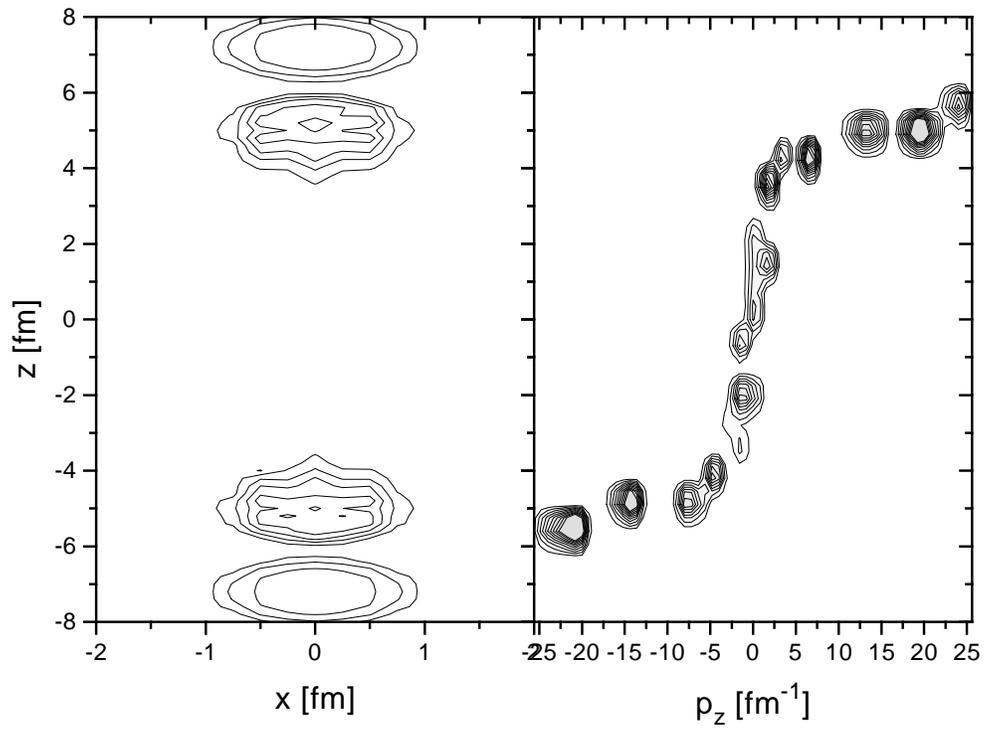
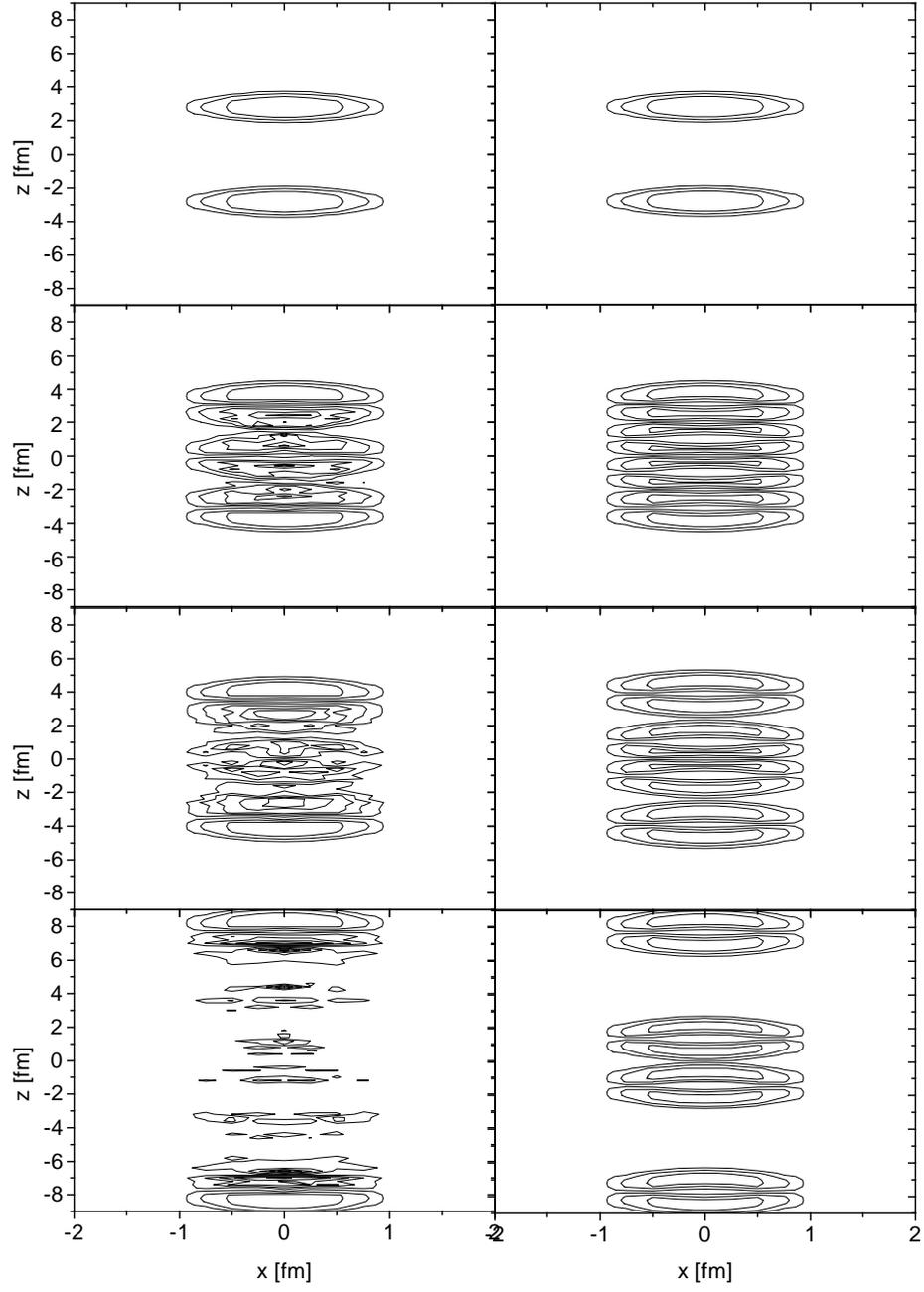
\end{document}